\documentstyle[aps,epsfig,twocolumn]{revtex}
\begin{document}
\twocolumn[\hsize\textwidth\columnwidth\hsize\csname @twocolumnfalse\endcsname
\draft
\title{
Metal-Insulator Transition Accompanied with a Charge Ordering in
the One-dimensional $t-J'$ Model
}
\author{Tetsuya Mutou, Naokazu Shibata and Kazuo Ueda}
\address{Institute for Solid State Physics, University of Tokyo,\\
7-22-1 Roppongi, Minato-ku, Tokyo 106, Japan
}
\date{\today}

\maketitle

\begin{abstract}
We study the metal-insulator transition accompanied with 
a charge ordering in the one-dimensional
(1D) $t-J'$ model at quarter filling by the density matrix 
renormalization group method. 
In this model the nearest-neighbor hopping energy $t$ competes
with the next-nearest-neighbor exchange energy $J'$.
We have found that a metal-insulator transition occurs at a
finite value of $t/J'$; $(t/J')_{\rm C} \simeq 0.18$ and the 
transition is of first order.
In the insulating phase for small $t/J'$, 
there is an alternating charge ordering 
and the system behaves as 
a 1D quantum Heisenberg antiferromagnet.
The metallic side belongs to 
the universality class of the Tomonaga-Luttinger liquids.
The quantum phase transition is an example of melting of 
the 1D quantum Heisenberg antiferromagnet.

\end{abstract}

\vskip2pc]

\narrowtext

Recently low-dimensional copper oxides  
have attracted much attention as a route to understand the physics of
the high-$T_{\rm C}$ superconductivity. 
Since the superconductivity in a doped ladder system
with a spin gap was theoretically suggested \cite{LadSuper}, 
several experimental 
studies of the quasi one-dimensional (1D) systems with 
the ladder structure have been carried out. 
Especially, the discovery of superconductivity in the 
quasi 1D material Sr$_{x}$Ca$_{14-x}$Cu$_{24}$O$_{41.84}$
($x=0.4$) under a high pressure 
has received considerable interest \cite{Uehara}.
The parent material Sr$_{14}$Cu$_{24}$O$_{41}$ 
includes CuO$_{2}$-chain layers and 
Cu$_{2}$O$_{3}$-ladder layers separated by the Sr layers \cite{Sr14}. 
Spin excitation gaps are observed in both chains and ladders 
in ambient pressure \cite{Sr14-Res,spingap,MatsudaESR}. 
The origin of the spin gap in the chain is still controversial
in spite of intensive studies 
\cite{MatsudaESR,Matsuda96,Hiroi96,Motoyama97,Eccleston97,Takigawa97}.
By neutron scattering measurements, Matsuda {\it et al.} concluded
\cite{Matsuda96}
that the spin gap in the chain originates from dimer
formation between the copper pairs separated by
two and four distances of the nearest neighbor copper
ions. On the other hand, in more recent 
neutron scattering measurements \cite{Eccleston97}, 
Eccleston {\it et al.} suggested that dimers are 
formed between next-nearest-neighbor Cu ions which are 
arranged with a period of 5 Cu-Cu distances in the chain \cite{Takigawa97}.
In the both models considered by Eccleston {\it et al.} and Matsuda 
{\it et al.}, the superexchange interactions between the 
next-nearest-neighbor Cu spins play an important role.

Matsuda {\it et al.}'s results have been interpreted in
\cite{Motoyama97}
that there exists the alternating Cu$^{+2}$ and Cu$^{+3}$
order along the chain. According to this
scenario, the mean valence of Cu ion in the chain is Cu$^{+2.5}$. 
On the contrary, Cu$^{+2.6}$ is assumed in 
Eccleston {\it et al.}'s results \cite{valence2.6}. 
Which scenario is realized is still an open problem. 
Note that for either case of Cu$^{+2.5}$ or Cu$^{+2.6}$,
the Cu $3d$ band is partially filled;
the highest $3d$ band is nearly quarter-filled.
Although the Cu $3d$ band is partially 
filled, the electrical resistivity of the system shows a
semiconducting temperature dependence \cite{Sr14-Res}. 
Precise understanding of the 
semiconducting temperature dependence depends on whether the system
is just at the quarter filling or not, since localization effects
should be taken into account for the latter case.

In the edge-sharing CuO$_{2}$ chain, the superexchange interaction
between next-nearest-neighbor Cu spins through Cu-O-O-Cu path 
is expected to be more important than 
the nearest-neighbor superexchange
interaction, 
since in the latter case the bond angle of Cu-O-Cu is close 
to 90 degree.
Thus, the model we consider in the present study is 
the simplest case where
the next-nearest-neighbor superexchange interaction competes 
with the kinetic energy.
For the kinetic energy we consider only the nearest-neighbor hoppings,
since in the vicinity of the metal-insulator transition of 
a quarter filled system the nearest-neighbor hoppings are more 
important than the next-nearest-neighbor hoppings even if the matrix
element of the latter processes is bigger than the former.
We assume strong Coulomb repulsion between the $d$-electrons,
leading to suppression of the double occupancy of the carriers.
This model may be called as the one-dimensional $t-J'$ model
where $t$ represents the nearest-neighbor hoppings and $J'$ 
the next-nearest-neighbor exchange.

\begin{figure}
\begin{center}
\leavevmode
\epsfxsize=45mm \epsffile{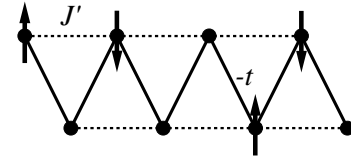}
\caption{
The $t-J'$ chain model. Each dot denotes a site and
arrows represent carriers with spins specified by arrows.
}
\label{fig:model}
\end{center}
\end{figure}
The model Hamiltonian is defined as
\begin{equation}
{\cal H} = -t\sum_{i\sigma}(
\tilde{c}^{\dag}_{i \sigma}\tilde{c}_{i +1 \sigma} + {\rm H.c.})
+J'\sum_{i}\mbox{\boldmath $S$}_{i}\cdot\mbox{\boldmath $S$}_{i +2},
\label{eqn:Ham}
\end{equation}   
where the operator $\tilde{c}^{\dag}_{i \sigma}$ 
($\tilde{c}_{i \sigma}$) creates (annihilates) 
a particle with a spin $\sigma$ at $i$th site.
These operators are defined in the subspace where  
double occupancy is excluded.
$\mbox{\boldmath  $S$}_{i}$ denotes 
the spin operator at $i$th site.
The model is topologically equivalent to the zigzag chain
which is shown in Fig.\ \ref{fig:model}. 
The solid and broken lines denote 
the nearest-neighbor hoppings $t$ and the next-nearest-neighbor
exchange interactions $J'$, respectively. 

In this paper we restrict ourselves to the quarter filling. 
At this filling, the system has two simple limits.
(i) $t \rightarrow 0 $, $J'$ is finite :
In this limit, 
all particles localize at odd or even sites, and 
the system has a gap in the charge excitations.
This charge-ordered state behaves as an antiferromagnetic (AF)
Heisenberg chain with the exchange interaction $J'$ \cite{Strict}. 
In the AF Heisenberg chain, it is well known that 
there is no gap in the spin excitations.
(ii) $J' \rightarrow 0$, $t$ is finite :
In the case of $J'=0$, the system is identical to the spinless fermions 
with the macroscopic spin degeneracy of $2^{N}$ ($N$ is the number of 
spins).  
For an infinitesimal $J'$
the spin degeneracy is lifted and the system is expected to 
belong to the universality class of the Tomonaga-Luttinger (TL)
liquids \cite{TLliquid}. In this limit,
the system is gapless in both the charge and spin excitations.
Since the present system has the two limits mentioned above, 
one can expect that a quantum phase transition occurs 
at some critical value  
$(t/J')_{\rm C}$ from an insulating phase (the charge gap
$\Delta_{\rm c}$ is finite) to a metallic phase 
($\Delta_{\rm c} = 0$) as increasing $t/J'$ from $0$ to $\infty$.
Since particles are expected to order alternatively
for a smaller $t/J'$ than $(t/J')_{\rm C}$, 
the quantum phase transition is a charge ordering transition.

The purpose of the present study is to answer the following two questions: 
(i) Does the quantum phase 
transition occur at a finite $(t/J')_{\rm C}$ or already at
infinitesimal $t/J'$ ?  
(ii) What is the nature of the phase transition?  
Is it a first- or second-order transition?

In order to determine the transition point, we calculate two 
quantities:
the charge density order parameter $n_{\rm cd}$ and 
the charge gap $\Delta_{\rm c}$.
In the insulating
phase, it is expected that the all particles localize at even or
odd sites. Therefore, 
we consider as the order parameter $n_{\rm cd} =
n_{\rm odd}-n_{\rm even}$, where $n_{\rm odd} (n_{\rm even})$ 
is the mean value of the number density $n_{i}$ 
at odd (even) sites. 
The order parameter $n_{\rm cd}$ is defined as
\begin{equation}
n_{\rm cd} \equiv \frac{\displaystyle 1}{\displaystyle L/2} 
\sum_{i = {\rm odd}}
(n_{i}-n_{i+1}).
\label{eqn:n-co}
\end{equation}
While $n_{\rm cd}$ should be zero in the metallic phase, 
it should be finite in the insulating phase.
In a finite-size system, the symmetry breaking does not occur 
if there is no external field. Thus, we calculate 
$n_{\rm cd}$ with a small external field $\epsilon$ 
at the end of the system.
This external field
is added to the model Hamiltonian (\ref{eqn:Ham}) as ${\cal H}-\epsilon
\sum_{\sigma}\tilde{c}_{1 \sigma}^{\dag}\tilde{c}_{1 \sigma}$ at the
first site ($\epsilon > 0$). 
We study the response of the system against the small external
field. 

The charge gap is defined by 
\begin{eqnarray}
2\Delta_{\rm c}(L) &\equiv&  
E_{\rm g}(N_{0}+2,0;L)+E_{\rm g}(N_{0}-2,0;L) \nonumber \\
& & -2E_{\rm g}(N_{0},0;L),
\label{eqn:chG}
\end{eqnarray}
where $E_{\rm g}(N,S_{z};L)$ denotes the ground state energy of
the system
in which the number of particles, the $z$ component of the 
total spin and the system size are denoted by $N$, $S_{z}$ and $L$,
respectively. In the quarter-filled case, $N=N_{0} \equiv L/2$.

\begin{figure}
\begin{center}
\leavevmode
\epsfxsize=65mm \epsffile{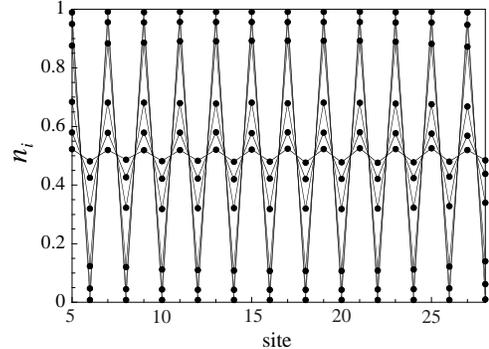}
\caption{
Local number density $n_{i}$ at each site for various values
of $t/J'$; $t/J' = 0.05$, $0.10$, $0.13$, $0.14$, $0.15$ and 
$0.17$. For a small $t/J'$, 
$n_{\rm odd} \simeq 1$ and $n_{\rm even} \simeq 0$.
Ths system size $L$ of this example is $32$ 
($\epsilon/J' = 1.0 \times 10^{-3}$).
}
\label{fig:localDNS}
\end{center}
\end{figure}
To study the quantum phase transition numerically,
it is essential to look at the size dependence systematically 
and therefore we need to treat large-size systems. 
To this end, we use the density matrix
renormalization group (DMRG) method \cite{DMRG}, which is one of the 
standard numerical methods to study one-dimensional quantum systems.
We study systems of various sizes ($L= 12, 16, 24, 32$, and $48$) 
by using the finite-system algorithm of the DMRG method \cite{DMRG}. 
We keep $100 \sim 150$ states in the renormalization procedures 
so that truncation errors in the density matrix are less than
$1.0 \times 10^{-4}$ (typical numerical errors are about 
$1.0 \times 10^{-5}$).

\begin{figure}
\begin{center}
\leavevmode
\epsfxsize=65mm \epsffile{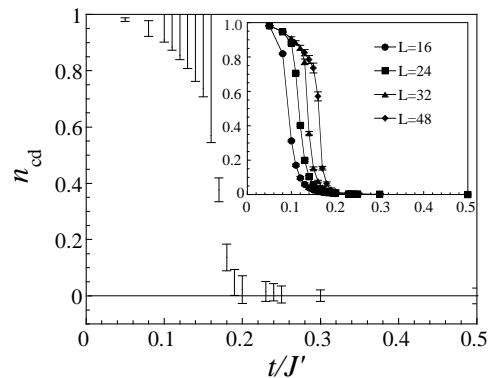}
\caption{
Results of size extrapolation of $n_{\rm cd}$ vs $t/J'$ (see the text).
In the inset we plot $n_{\rm cd}$ for various system sizes.
}
\label{fig:nD-size}
\end{center}
\end{figure}
Figure\ \ref{fig:localDNS} shows examples of the local number 
density $n_{i}$ for various values of $t/J'$ with an external
field $\epsilon/J' = 1.0 \times 10^{-3}$ ($L=32$). 
Shown here are 
central 28 sites excluding 4 sites at both ends of the system.
For small $t/J' (\ll 1)$, 
one can clearly see charge ordering behaviors.
Most particles spontaneously localize at odd or even sites 
and in this example the odd sites are selected by the small 
symmetry breaking field at the first site.
For a small $t/J'$ charge fluctuations are small, i.e., 
$n_{\rm odd} \simeq 1$ and $n_{\rm even} \simeq 0$.
As increasing the value of $t/J'$,
the difference between $n_{\rm odd}$ and $n_{\rm even}$ decreases
for a fixed system size.
Both $n_{\rm odd}$ and $n_{\rm even}$ approach the mean value
of the number density $N_{0}/L=0.5$ for large $t/J'$.

Charge density order parameters $n_{\rm cd}$ are shown  
in the inset of Fig.\ \ref{fig:nD-size} 
for various system sizes ($L=16, 24, 32$, and $48$). 
Error bars in the inset represent the standard deviation,
\begin{equation} 
\sqrt{\frac{\displaystyle 1}{\displaystyle L/2}
\sum_{i = {\rm odd}}(n_{i}-n_{i+1})^{2}
-n_{\rm cd}^{2}},
\label{eqn:Var}
\end{equation}
but for most of data they are smaller than the symbols. 
One can see that as the system size is increased, 
the disappearance of $n_{\rm cd}$ 
becomes sharper.
The region of the charge ordered phase 
spreads out for larger-size systems.
On the other hand, for all system sizes 
$n_{\rm cd}$ vanishes in the region
$t/J' \gtrsim 0.2$.
These results may be an indication that the charge ordering transition
is of the first order. However to draw a definite conclusion careful
finite size scaling is necessary.
Results of size extrapolation of $n_{\rm cd}$
are shown in the main part of Fig.\ \ref{fig:nD-size}.
We have carried out the linear extrapolation using data for 
the system size $L=24$ and $32$.
For the transition region $0.13 \le t/J' \le 0.20$, 
we have used data for $L=32$ and $48$.
In the case that the extrapolated value obtained 
by the simple linear extrapolation is greater than $1$, 
the upper error bar is fixed to unity. 
The values of $n_{\rm cd}$ for $t/J' \gtrsim 0.2$ are extrapolated 
to zero, while the values below $t/J' \simeq 0.2$ tend to be finite.
Concerning the symmetry breaking field,
we have confirmed that 
$n_{\rm cd}$ in the metallic side are suppressed even more 
and the disappearance of $n_{\rm cd}$ becomes sharper
for a smaller external field $\epsilon$
($=1.0 \times 10^{-4}J'$).

\begin{figure}
\begin{center}
\leavevmode
\epsfxsize=65mm \epsffile{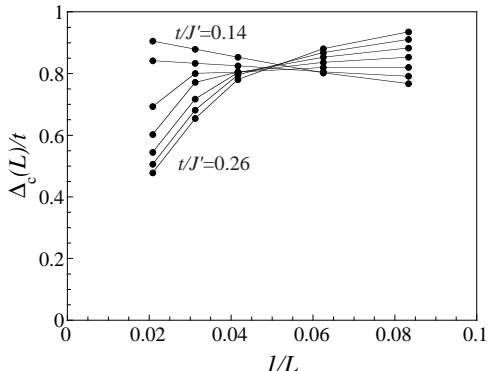}
\caption{
Size dependence of the charge gap $\Delta_{\rm c}(L)$ scaled by $t$ 
for several values of $t/J'$; $t/J'=0.14$, $0.16$, $0.18$, $0.20$,
$0.22$, $0.24$ and $0.26$.
Note that the system size dependence is qualitatively different
between $t/J'=0.14, 0.16$ and $t/J' \ge 0.18$. 
}
\label{fig:gap-c}
\end{center}
\end{figure}
Now we turn to the charge gap.
In Fig.\ \ref{fig:gap-c}, we present the system-size 
dependence of 
the charge gap scaled by $t$ for several values of $t/J'$. 
The system-size dependence of $\Delta_{\rm c}(L)$
for $t/J' \le 0.16$ is clearly different from that for 
$t/J' \ge 0.18$. 
For $t/J' \le 0.16$, $\Delta_{\rm c}(L)$ increases as $L$ is
increased, which suggests that the infinite-size limit
of $\Delta_{\rm c}$ is finite. One can conclude that 
the system is in the insulating
phase for $t/J' \le 0.16$.
On the other hand, for $t/J' \ge 0.18$, it tends to zero in the 
infinite-size limit; the system is in the metallic phase.
The observed behaviors of the charge gap and the system-size
dependence of the charge density order parameters are compatible
with the first-order phase transition at $(t/J')_{\rm C} \sim 0.18$.

\begin{figure}
\begin{center}
\leavevmode
\epsfxsize=65mm \epsffile{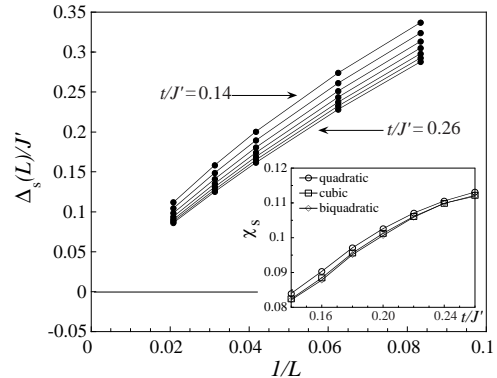}
\caption{
Size dependence of the spin gap $\Delta_{\rm s}(L)$ scaled by $J'$ 
for several values of $t/J'$; $t/J'=0.14$, $0.16$, $0.18$, $0.20$,
$0.22$, $0.24$ and $0.26$ (from top to bottom).
The inset shows $\chi_{\rm s}$ evaluated 
from $\Delta_{\rm s}(L)$ (see the text).
}
\label{fig:gap-s}
\end{center}
\end{figure}
We also calculate the spin gap defined as 
\begin{equation}
\Delta_{\rm s}(L) 
\equiv E_{\rm g}(N_{0},1;L)-E_{\rm g}(N_{0},0;L),
\label{eqn:spG}
\end{equation}
in order to
see whether there is an anomaly in the spin excitations at
the quantum phase transition.
The system-size dependence of the spin gap scaled 
by $J'$ is shown in Fig.\ \ref{fig:gap-s}.
In both (insulating and metallic) phases, $\Delta_{\rm s}(L)$ tends 
to zero in the infinite-size limit and one can conclude that
there is no gap in the spin excitations. 
The slope 
${\rm d}\Delta_{\rm s}(L)/{\rm d}(1/L)|_{(1/L)=0}$
gives the spin susceptibility $\chi_{\rm s}$ of the system through
the relation $\Delta_{\rm s}(L)=2/(\chi_{\rm s}L)$ \cite{Vs}.
The coefficient of the linear term is determined by a fitting to a
quadratic function including 
the origin;($1/L=0$, $\Delta_{\rm s}(L=\infty)=0$), 
points for $L=48$; ($1/48$,$\Delta_{\rm s}(48)$)
and for $L=32$; ($1/32$,$\Delta_{\rm s}(32)$).
We obtain the coefficients also by using cubic and
biquadratic functions including four (the origin, points for 
$L=48$, $32$, and $24$) and five (the origin, points for $L=48$, $32$,
$24$, and $16$) points, respectively.
The inset of Fig.\ \ref{fig:gap-s} shows $\chi_{\rm s}$ vs $t/J'$
obtained by the above-mentioned procedures.
Good coincidence between the results 
evaluated from cubic and biquadratic functions indicates that 
the values thus obtained for $\chi_{\rm s}$ are reliable.
One can see that $t/J'$-dependence of $\chi_{\rm s}$ for $t/J' < 0.18$
may be slightly different from that for $t/J' \ge 0.18$, although it is not
possible to conclude whether there is a discontinuity 
at $(t/J')_{\rm C} \simeq 0.18$ or not.
Since the phase transition occurs at $J'$ 
which is about five times bigger than $t$, it is natural to assume
that the energy scale of system in the region near the critical point 
is dominated by $J'$. A natural consequence is that the anomaly 
at the critical point, if any, should be small.

\begin{figure}
\begin{center}
\leavevmode
\epsfxsize=60mm \epsffile{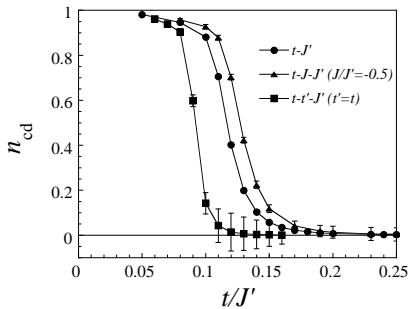}
\caption{
Comparison of $n_{\rm cd}$ in the $t-J-J'$ chain ($J/J'=-0.5$)
and $t-t'-J'$ chain ($t'=t$) with that in the $t-J'$
chain. The system size is $L=32$ ($\epsilon/J' = 1.0 \times 10^{-3}$).
}
\label{fig:comparison}
\end{center}
\end{figure}

The present model is simple but rather special in the sense that
the nearest-neighbor exchange energy $J$ and 
the next-nearest-neighbor hopping energy $t'$ are neglected 
when $t$ and $J'$ are considered. 
In spite of the special nature of the model,
the metal-insulator transition with the charge ordering is
generic. To show this, 
we investigate a few cases with the nearest-neighbor 
exchange or the next-nearest-neighbor hopping. In Fig.\ 
\ref{fig:comparison}, we show $n_{\rm cd}$ in the following two cases:
with an additional nearest-neighbor (ferromagnetic) exchange $J/J'=-0.5$ and
with a next-nearest-neighbor hopping $t'=t$.
In both cases, for small $t/J'$, $n_{\rm cd}$ is nearly equal to 1,
and it disappears rapidly around some critical value of $t/J'$.
These behaviors are qualitatively similar to that of the $t-J'$ chain 
model which we have discussed in the present paper in detail.
In the case with $J$ ($t-J-J'$ model), 
the critical value of $t/J'$ shifts to a 
larger value than that of the $t-J'$ chain model
since the hopping energy $t$ competes with 
not only $J'$ but also $J$. 
On the other hand, in the case
with $t'(=t)$ ($t-t'-J'$ model), 
the critical value of $t/J'$ is smaller than that of
the $t-J'$ chain model. This is natural since in the $t-t'-J'$ model 
the gain of the kinetic energy in the metallic phase 
is larger than that in the $t-J'$ model.

In conclusion, 
we have calculated the charge density order parameter,
the charge gap, and the spin gap obtained in the $t-J'$ chain model
using the DMRG method. We have found that  
the quantum phase transition from a charge ordered
insulator to a metal occurs at $t/J' \simeq 0.18$ at 
quarter filling,
and the transition is of first order. 
The transition is generic
in the models in which the kinetic energy competes with the
next-nearest-neighbor exchange energy at quarter filling.
For the present model, an interesting question is existence or absence
of the spin gap close to the metal-insulator transition.
The spin excitations are gapless  
in both limits $t \rightarrow 0$ and $J' \rightarrow 0$.
We have found that the spin excitations remain gapless over the whole 
$t/J'$ range between the metallic side and 
the insulating side. The spin susceptibility is obtained 
>from the spin gap of the finite systems.
One can see that 
the $t/J'$-dependence of $\chi_{\rm s}$ 
in the metallic phase is slightly 
different from that in the insulating phase, and there is a
possibility that a small anomaly exists at the critical point
in the bulk system. 

The quantum phase transition discussed in the present paper
is an example of melting of the 1D quantum Heisenberg 
antiferromagnet.
In this example, 
the melting is of first order and the metallic side belongs to 
the universality class of the TL liquids.

We are grateful to H. Kontani and M. Sigrist for helpful discussions,
and to H. Tsunetsugu for useful comments.
A part of the numerical calculations was done by SR2201 at the 
Center for Promotion of Computational Science and Engineering 
of Japan Atomic Energy Reserach Institute.
This work is financially supported by Grant-in-Aid from the 
Ministry of Education, Science and Culture of Japan.


\begin{thebibliography}{99}
\bibitem{valence2.6}
The valence Cu$^{+2.6}$ in the chain
is simply obtained if one assumes that the valence of Cu 
in the ladder is Cu$^{+2}$.
\bibitem{Strict}
Strictly speaking, in the case of $t=0$, there is 
the macroscopic degeneracy of configurations of particles in the
thermodynamic limit.
\bibitem{Vs} The spin susceptibility $\chi_{\rm s}$ is related to 
the spin velocity $v_{\rm s}$ 
for a TL liquid by the relation 
$\chi_{\rm s}=1/(2\pi v_{\rm s})$.
\end{thebibliography}
\end{document}